\documentclass[a4paper]{jpconf}
\usepackage{graphicx}
\usepackage{amssymb}
\usepackage{amsmath}

\newcommand{\be}{\begin{equation}} 
\newcommand{\ee}{\end{equation}} 
\newcommand{\ba}{\begin{eqnarray}} 
\newcommand{\ea}{\end{eqnarray}} 

\begin{document}
\title{Strange meson-baryon interaction in hot and dense medium: recent progress for a road to GSI/FAIR}

\author{D Cabrera$^{1,2}$, L Tolos$^{1,3}$, J Aichelin$^4$ and E Bratkovskaya$^{1,2}$}

\address{
$^1$ Frankfurt Institute for Avanced Studies, J.~W.~Goethe-Universit\"at, Frankfurt am Main, Germany\\
$^2$ Institut f\"ur Theoretische Physik, J.~W.~Goethe-Universit\"at, Frankfurt am Main, Germany\\
$^3$ Institut de Ci\`encies de l'Espai (IEEC/CSIC), Universitat Aut\`onoma de Barcelona, Cerdanyola del Vall\`es, Spain\\
$^4$ Subatech, UMR 6457, IN2P3/CNRS, \'Ecole de Mines de Nantes, Universit\'e de Nantes, France
}

\ead{cabrera@fias.uni-frankfurt.de}

\begin{abstract}
We report recent results on the dynamics of strange hadrons in two-body reactions relevant for near-threshold production in heavy-ion collisions at GSI/FAIR and NICA-Dubna. 
In particular, $\bar K N$ scattering in hot and dense nuclear matter is studied within a chiral unitary framework in coupled channels, setting up the starting point for implementations in microscopic off-shell transport
approaches. We focus on the calculation of transition rates with special attention to the excitation of hyperon resonances and isospin effects.
Additionally, we explore ``unconventional'' strangeness generation by meson-meson and meson-baryon interactions in connection with recent HADES observations of deep sub-threshold $\phi$ and $\Xi$ production.
\end{abstract}

Hadrons with strangeness, embedded in strongly interacting matter at finite temperature and density, have been matter of intense investigation in the last decades in connection with the study of exotic atoms \cite{Friedman:2007zza}, the analysis of strangeness production in heavy-ion collisions (HICs) \cite{Fuchs:2005zg,Forster:2007qk,Hartnack:2011cn} and the microscopic dynamics ruling the composition of neutron stars \cite{Kaplan:1986yq}.
In particular, understanding the dynamics of light strange mesons in a nuclear environment has posed a challenge to theoretical models. 
The $\Lambda(1405)$ resonance, sitting below the $\bar KN$ threshold in vacuum, is excited at a higher energy in nuclear matter due to Pauli-blocking on the nucleons. This leads to a rapid change of the $\bar KN$ scattering amplitude from repulsive in vacuum to attractive in the medium already at small densities, as required by kaonic-atom phenomenology \cite{Friedman:2007zza}, quickly deviating from the low-density approximation.
The production of $K$ and $\bar K$ close to threshold energy has also been thoroughly investigated in HICs at SIS energies \cite{Fuchs:2005zg,Forster:2007qk,Hartnack:2011cn,Cassing:2003vz,Tolos:2003qj}. The analysis of experimental data in conjunction with microscopic transport approaches has allowed to draw solid conclusions regarding the production mechanisms  and the freeze-out conditions of strange mesons. Still, a simultaneous description of all observables involving antikaon production is missing \cite{Hartnack:2011cn}. For example, recent experimental data on the $v_1$, $v_2$ flow coefficients of strange mesons show sensitivity to the details of the in-medium meson-baryon interaction, leaving room for improvements within hadronic models accounting for medium effects.

The properties of strange mesons in nuclear matter at finite temperature were studied within a self-consistent coupled-channel approach based on the SU(3) meson-baryon chiral Lagrangian (see \cite{Tolos:2006ny,Tolos:2008di} and references therein). Such kind of approaches successfully describes the $\bar K$ meson interactions in nuclear matter, accounting for a shallow attractive $\bar K$ potential at nuclear matter density ($40$-$60$~MeV) and a broad spectral function. Our present results improve on this model by implementing the unitarization of $\bar K N$ scattering amplitudes in both $s$ and $p$ waves at finite density and temperature \cite{Cabrera:2014lca}.
This formulation gives access to full off-shell in-medium scattering amplitudes in the SU(3) set of coupled channels. For $K^- p$ scattering these include: $K^- p$, $\bar{K}^0n$, $\pi^0
\Lambda$, $\pi^0 \Sigma^0$, $\eta \Lambda$, $\eta \Sigma^0$,  $\pi^+
\Sigma^-$, $\pi^- \Sigma^+$, $K^+ \Xi^-$ and $K^0 \Xi^0$; whereas for $K^- n$ scattering we have $K^- n$, $\pi^0 \Sigma^-$, $\pi^- \Sigma^0$, $\pi^- \Lambda$, $\eta \Sigma^-$ and $K^0 \Xi^-$ [compare to, e.g., the previous analysis \cite{Cassing:2003vz} within the Hadron-String-Dynamics (HSD) transport model].
Another important asset of our approach is that, whereas it achieves a good description of $K^- p$ vacuum low-energy scattering observables with a minimal number of parameters, the model is highly predictive also in the $K^- n$ sector. The relevance of isospin effects in strangeness production near threshold was pointed out in \cite{Hartnack:2011cn}.

\begin{figure*}
\centering
\includegraphics[height=5cm,width=0.45\textwidth,clip]{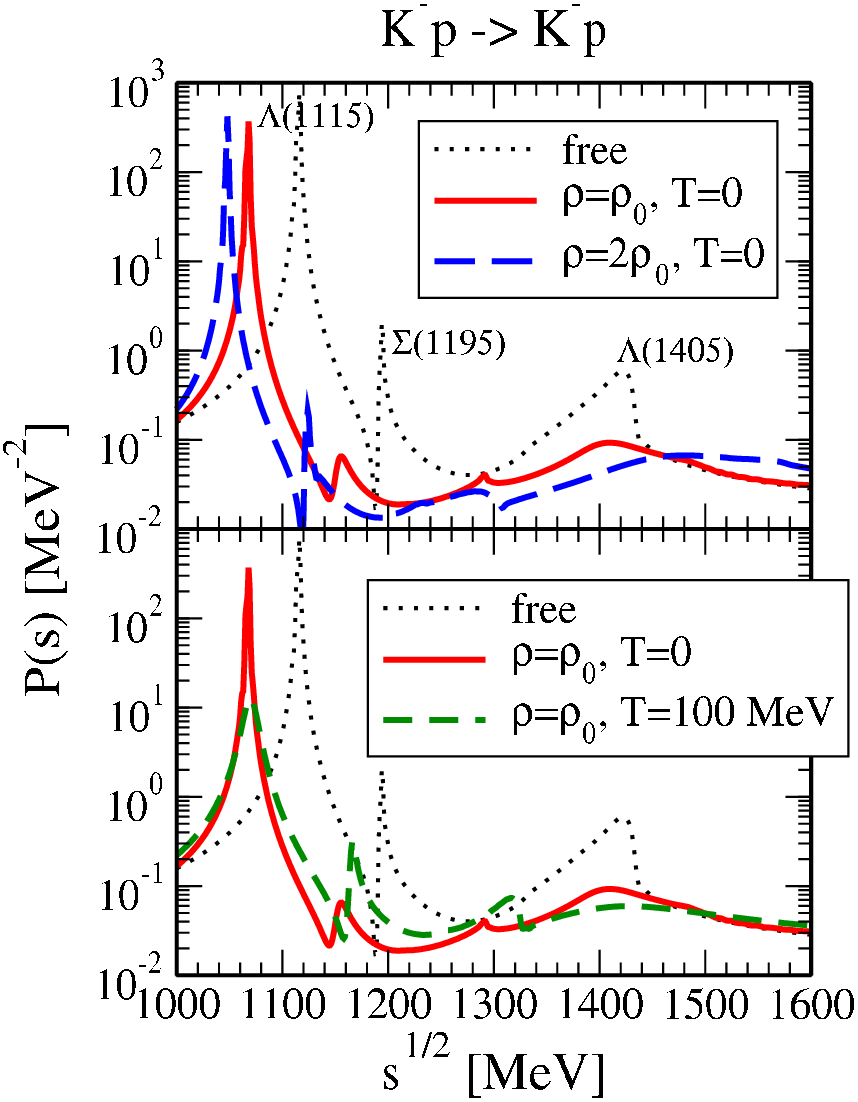} %was 12cm
\includegraphics[height=5cm,width=0.48\textwidth,clip]{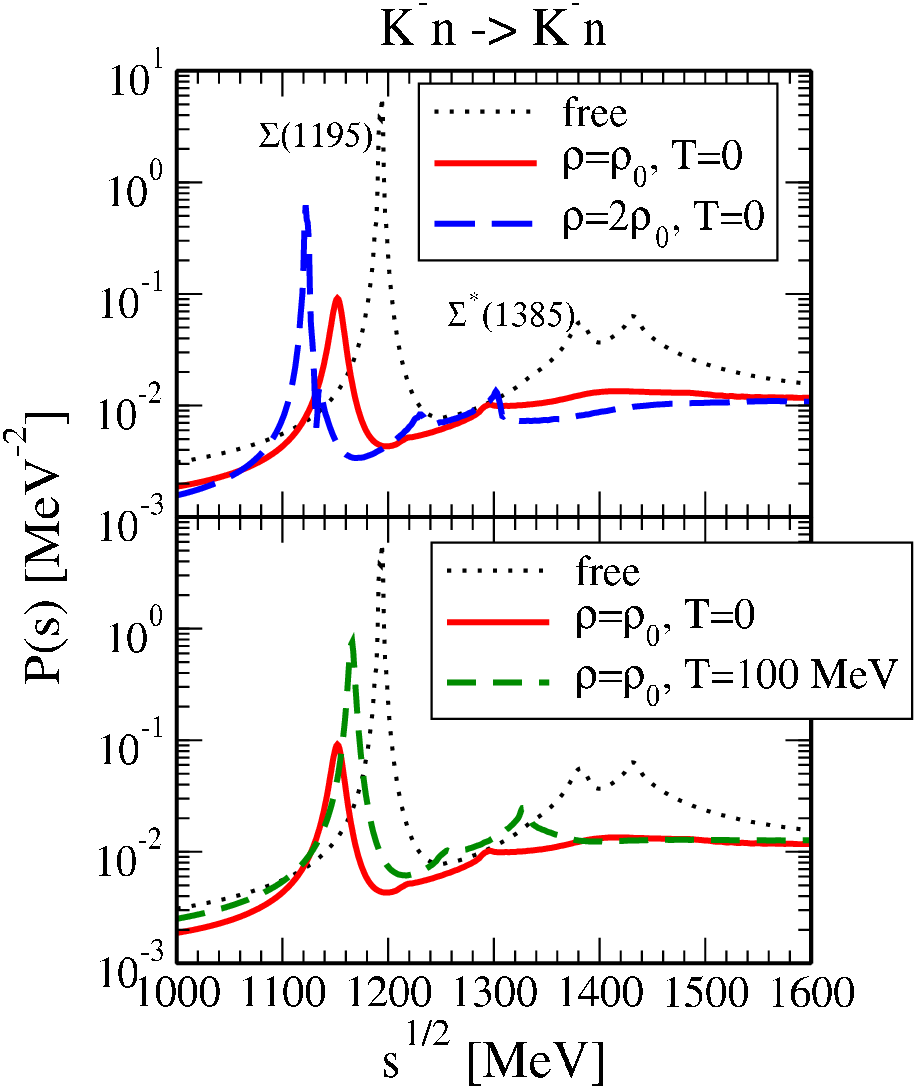} %was 12cm
\caption{In-medium transition probability ${\cal P}$ at zero total three-momentum of the meson-baryon pair for the elastic $K^-p$ (left) and $K^- n$ (right) reactions.}
\label{fig:kmp-vs-kmn}       % Give a unique label
\end{figure*}

The calculation of dynamical quantities in transport theory requires that the pertinent reaction rates or transition probabilities are folded with the spectral functions of the particles in the initial and final states \cite{Cassing:2003vz}.
The transition probability for a given reaction, ${\cal P}$, is defined fully off-shell as the angular integrated modulus-squared scattering amplitude (including all partial waves, $L=0,1$), averaged over the total angular momentum ($J=1/2, 3/2$). The cross section for the process $i\to j$ then follows as $\sigma_{ij} = \frac{1}{8 \pi} \frac{M_i M_j}{s} \frac{\tilde{q}_j}{\tilde{q}_i}\,{\cal P}_{ij}$, with $\tilde{q}_i$ the c.m. three-momentum and $M_i$ the baryon mass in channel $i$. The total cross section for a given reaction (e.g.~$K^-p$) is then given by the sum over the partial cross sections for all the permitted coupled channels, $\sigma^i_{\rm tot} = \sum_j \sigma_{ij}$.
We emphasize here that in our model we also have access to the angular dependence of the scattering amplitudes, which is non-trivial due to the contribution of the (deeply) sub-threshold hyperon excitations in the $p$-wave ($\Lambda$, $\Sigma$ and $\Sigma^*$). The angular information is a novel aspect of our study and allows for a full determination of the kinematics of the outgoing particles.

%%%%%% Results: transition rates
In Fig.~\ref{fig:kmp-vs-kmn} the transition probability for the $K^-p$ and $K^- n$ elastic reactions is depicted as a function of the meson-baryon center-of-mass energy at total vanishing three-momentum, evidencing the importance of isospin effects.
The $K^-p$ state is an admixture of isospin $I=0,1$ and therefore the two isoscalar $\Lambda$ resonances and the isovector $\Sigma(1195)$ show up in the spectrum.
The $\Sigma^*(1385)$ couples weakly to the $\bar K N$ system and cannot be resolved in the $K^-p$ elastic reaction.
The $K^-n$ reaction is pure $I=1$ and consequently only the isovector hyperon excitations are present in the 
right panel of Fig.~\ref{fig:kmp-vs-kmn}. This explains the dramatic difference between the $K^-p$ and $K^- n$ cross sections in vacuum, the former being dominated by a quick fall due to a strong Flatt\'e effect on the resonance shape of the $\Lambda(1405)$.
In the medium,  the structure of the $\Lambda(1405)$ is practically washed out at normal matter density. The $p$-wave ground states experience moderately attractive mass shifts as density is increased, whereas the $\Sigma^*(1385)$ is largely broadened due to the opening of in-medium decay channels such as $\Sigma^*\to\Lambda NN^{-1}$, $\Sigma NN^{-1}$ (cf.~Fig.~5 in \cite{Cabrera:2014lca}). The effect of temperature is particularly appreciable as a broadening of the $p$-wave resonances as compared to the vacuum case.
Our results are complementary to the ones obtained in our previous work \cite{Tolos:2008di}, where the $\bar K$ spectral function and nuclear optical potential were provided. Altogether, they permit a systematic accounting of medium effects in the $S=-1$ sector within transport approaches such as PHSD and IQMD.

%\section{$\phi$ and $\Xi$ production in the hadronic phase}
The implementation of suitable hadronic models in PHSD offers a unique opportunity to address some of the standing puzzles regarding the production of strange mesons and baryons, such as the $\phi$ and the $\Xi$.
The observation of deep sub-threshold $\phi$ and $\Xi$ production by the HADES Collaboration in $1.23$~AGeV Au$+$Au collisions \cite{Lorenz:2014eja} calls for a review of mechanisms that may have been missed in the analysis performed by different transport simulation groups, involving the production from resonance decay \cite{Steinheimer:2015sha} as well as from two-body hadronic processes \cite{Pierre-SQMproc}.
We briefly report here on recent progress on the theoretical assessment of pertinent meson-meson and meson-baryon reactions in the light of chiral approaches.

A suitable candidate is the $\eta \pi \to \phi \pi$ reaction as the advocated hidden strangeness content of the $\eta$ could help circumvent OZI suppression in meson-meson interactions as compared to $\phi$ production in $NN$ collisions. Moreover, the $\eta \pi$ system couples strongly to the isovector $a_0(980)$, which may enhance the cross section as it sits not far below the $\phi \pi$ threshold with a peculiar cusp-like energy dependence. Unfortunately, the mismatch between initial and final states regarding $C$-parity ($G$-parity) in the case of $\eta \pi^0 \to \phi \pi^0$ ($\eta \pi^\pm \to \phi \pi^\pm$) suppresses this reaction and the relevant chiral amplitudes vanish at leading order (LO)\footnote{The small branching ratio of $\phi \to \pi^+ \pi^- \eta$ as compared to the dominant hadronic decay of the $\phi$ is a clear indication of this suppression.}. 

Next we have explored reactions involving the interaction of the $\eta$ with nucleons and deltas to produce a $\phi$ in the final state, e.g.~$\eta N(\Delta) \to \phi N(\Delta)$. Here the challenge dwells on the theoretical side, since a model for this process has to couple meson-baryon states with pseudoscalar and vector mesons from the SU(3) octet and baryons from the $1/2^+$ octet and the $3/2^+$ decuplet.
The low- and intermediate-energy interactions of pseudoscalar and vector mesons with baryons have been described successfully in coupled-channel unitarized approaches based on chiral symmetry (see, e.g., \cite{Oset:2012ap} and references therein). However, vector and pseudoscalar mesons are decoupled in these approaches whereas there is no reason, based on symmetry or dynamics, to expect such phenomenon.
With our goal in mind we recourse to the approach in Ref.~\cite{Gamermann:2011mq}, which implements an extension of the Weinberg-Tomozawa interaction of standard SU(3) meson-baryon ChPT by means of SU(6) spin-flavor symmetry. 
The model comprehends several sectors of total spin, isospin and strangeness (with negative parity), where a number of baryonic resonances are dynamically generated in the energy range from the $N^*(1535)$ up to $\sim$2.2~GeV from the unitarization of the LO scattering amplitudes. In particular we note that two resonances slightly above the $\phi N$ threshold emerge in this approach, with $J^P=1/2^-,3/2^-$. Such states, which also couple strongly to $K^*\Lambda$ and $K^*\Sigma$, have been found crucial to correctly interpret the data shape and energy dependence of the $\gamma p\to K^0\Sigma^+$ reaction by the CBELSA/TAPS Collaboration \cite{Ramos:2013wua}. Claims for the need of a $N^*$ resonant state in that energy region have also been made from the analysis of $\Lambda(1520)$ photoproduction data by the LEPS Collaboration.

\begin{figure*}
\centering
\includegraphics[width=0.48\textwidth,clip]{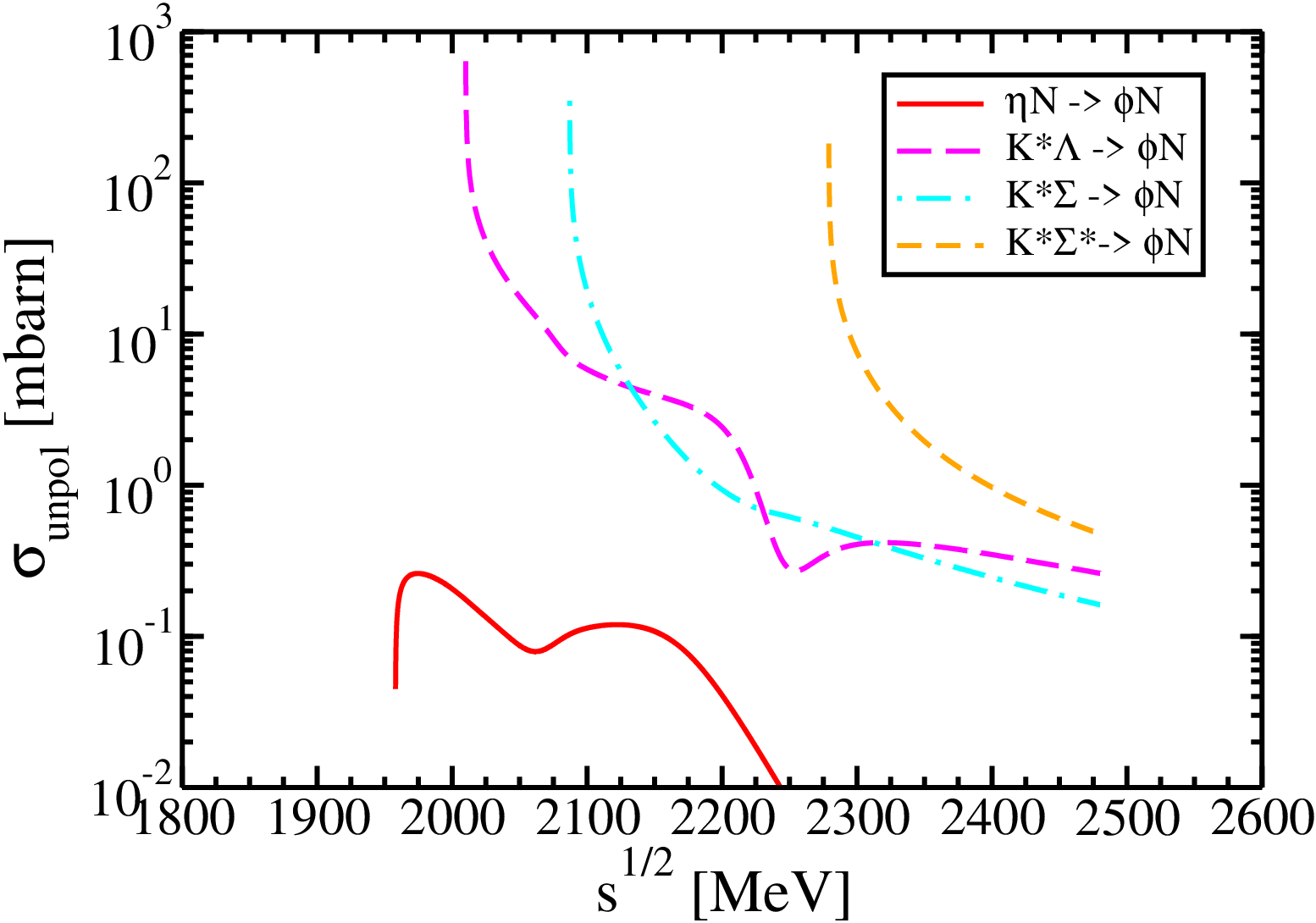} %was 12cm
\includegraphics[width=0.48\textwidth,clip]{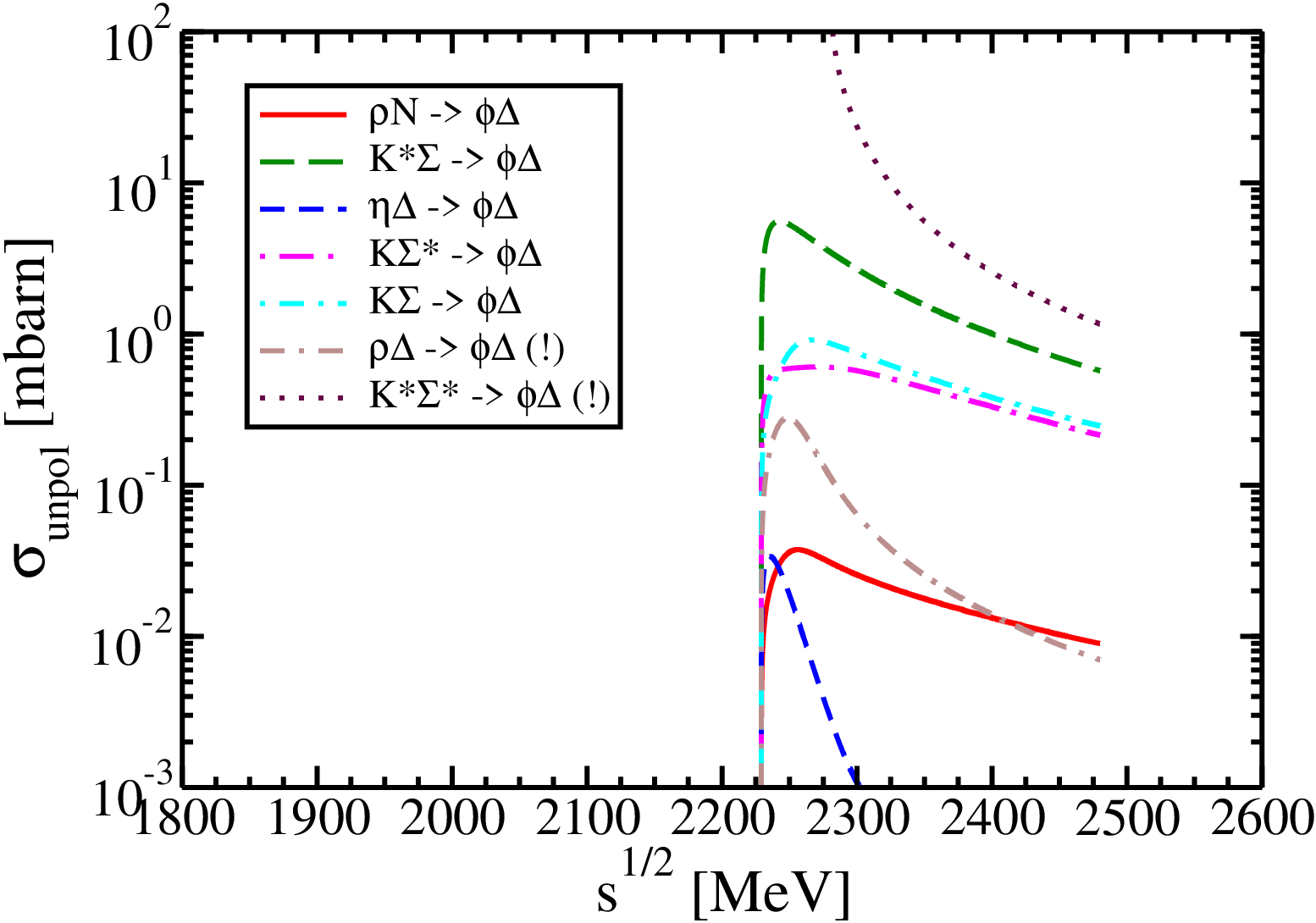} %was 12cm
\caption{Unpolarized cross section for $\phi$ production from several isospin $1/2$ (left) and $3/2$ (right) meson-baryon reactions. The (!) sign indicates processes involving up to $J=5/2$.}
\label{fig:phiprod}       % Give a unique label
\end{figure*}

We focus on the relevant channels producing a $\phi$ meson in the final state with total strangeness $S=0$ whereas other sectors (involving multi-strange hadrons, such as the $\Xi$) will be studied in the future.
In Fig.~\ref{fig:phiprod} we depict the cross section, as a function of the center-of-mass energy, for the $\eta N \to \phi N$ reaction (total isospin $I=1/2$). In spite of the resonant states present above $2$~GeV the cross section remains below the mbarn scale. The reason for this is twofold: First, the LO chiral amplitude for this process is vanishing, the coupling proceeding via an intermediate $K Y$ state. Second, whereas the states above 2~GeV have a large coupling to the final state, they are practically decoupled from $\eta N$, which lies far below in energy.
%The influence of resonances on the cross section can be gauged in terms of their coupling to the initial and final states, which can be estimated from the residue of the $T$-matrix at the resonance pole.
Motivated by the recent experimental findings above, we have also calculated the cross sections for $K^*Y \to \phi N$, with $Y=\Lambda,\Sigma,\Sigma^*$. Although the $K^*(892)$ is produced as a secondary particle, preliminary simulations in PHSD indicate that it could be sufficiently abundant in the hadronic phase as to influence $\phi$ production at least at SPS energies and beyond. Our results show sizable cross sections in the range of $10$-$100$~mbarn close to threshold energies, exhibiting the typical rapid drop of strangeness-exchange reactions and signalling the presence of nearby resonant states. Other reactions without net strangeness content have also been evaluated resulting in smaller cross sections, typically below $5$~mbarn [e.g.~$\rho N(\Delta)\to\phi N$]. Production in the $I=3/2$ sector (i.e.~with $\phi\Delta$ as final state) is also found to be relevant, as is shown if the right panel of Fig.~\ref{fig:phiprod}.
Further conclusions on present and future measurements of $\phi$ and $\Xi$ production
can be drawn from detailed studies within PHSD. Such developments are in progress.

\vspace{-0.2cm}
\section*{Acknowledgments}
We acknowledge stimulating discussions with W.~Cassing, E.~Oset and H.~Stroebele, and support
from HIC for FAIR, from BMBF under Project No. 05P12RFFCQ, from Grants No.~FPA2010-16963 (Ministerio de Ciencia e Innovaci\'on), No.~FP7-PEOPLE-2011-CIG under Contract No.~PCIG09-GA-2011-291679, No.~FPA2013-43425-P, and the EC-RII HadronPhysics3 (Grant Agreement No. 283286). L.T. acknowledges support from the Ram\'on y Cajal Research Programme (Ministerio de Ciencia e Innovaci\'on).

\end{document}